\documentclass[prl, twocolumn, showpacs, amssymb, amsmath,aps] {revtex4}

\usepackage{graphicx}% Include figure files
\usepackage{dcolumn}% Align table columns on decimal point
\usepackage{bm}% bold math
\usepackage{epstopdf}
\usepackage{latexsym}
\begin{document}

%\preprint{PRL/Clock}

\title{A Continuous Non-demolition Measurement of the Cs Clock Transition Pseudo-spin}

\author{Souma Chaudhury}
\author{Greg A. Smith}
\author{Kevin Schulz}
\author{Poul S. Jessen}
\affiliation{%
College of Optical Sciences, University of Arizona, Tucson, AZ 85721
}%

\date{\today}% It is always \today

\begin{abstract}
We demonstrate a weak continuous measurement of the pseudo-spin associated with the clock transition in a sample of Cs atoms. Our scheme uses an optical probe tuned near the $\mathrm{D_{1}}$ transition to measure the sample birefringence, which depends on the $z$-component of the collective pseudo-spin. At certain probe frequencies the differential light shift of the clock states vanishes and the measurement is non-perturbing. In dense samples the measurement can be used to squeeze the collective clock pseudo-spin, and has potential to improve the performance of atomic clocks and interferometers.\end{abstract}

% PACS, the Physics and Astronomy
                             % Classification Scheme.
                              
\maketitle

The design of highly sensitive measurements on a quantum manybody system is of interest both for applications and for the fundamental insight provided into the process of quantum measurement. For example, sensitive interrogation of atomic ensembles is important for metrology and sensing, most notably atomic clocks and atom interferometer-based inertial sensors. These devices are now limited by quantum fluctuations (projection noise) in the number of atoms measured in each clock state \cite{a} or interferometer output \cite{b}. For an uncorrelated $N$-atom ensemble the measurement signal and projection noise scales as $N$ and $\sqrt{N}$ respectively, leading to a measurement precision that increases as $\sqrt{N}$. As for optical interferometry \cite{c}, however, an appropriate entangled input state can suppress noise and allow measurement performance to approach the Heisenberg limit where the precision scales as $N$. Wineland and coworkers proposed the use of squeezed states of the pseudo-spin associated with a clock transition \cite{squeezeprop}, and demonstrated improved measurement precision in an ensemble of trapped ions \cite{d}. Others have explored the generation of similar squeezed states in large ensembles of alkali atoms, either by direct atom-atom interaction \cite{e} or by coupling to a shared mode of an optical cavity \cite{f}. Interestingly, the existence of projection noise limited measurements suggests a different approach: if a measurement can resolve the quantum fluctuations associated with a collective observable, then backaction will be induced on the collective state and the observable can be squeezed. This idea was explored in experiments that used the linear Faraday effect to perform a quantum non-demolition (QND) measurement of the collective hyperfine spin-angular momentum in an ensemble of alkali atoms \cite{g,h}. 

In this Letter we demonstrate a weak, continuous measurement of the clock pseudo-spin in a laser cooled Cs ensemble, with performance comparable to Faraday rotation-based measurements of hyperfine angular momenta \cite{faraday}. Our scheme employs a probe beam tuned in-between the $F=4 \to F'=3,4$  transitions of the $\mathrm{D_{1}}$ multiplet at 895 nm, where the sample is birefringent by an amount proportional to the population of the upper clock state. Measuring the induced probe ellipticity allows us to observe microwave driven Rabi oscillations between the clock states in real time and with good signal-to-noise ratio (SNR). We further show that the coherent part of the atom-probe coupling does not perturb the clock pseudo-spin or couple it to the rest of the ground hyperfine manifold, and that the useful measurement time is therefore limited by optical pumping. This suggests that measurement-based squeezing of the clock pseudo-spin will be possible for atomic samples that are optically dense on resonance, in close analogy to the experiment of Geremia et al. \cite{h}. A measurement of the clock pseudo-spin was previously implemented in \cite{i}, where the atomic sample was inserted in one arm of a Mach-Zehnder interferometer.  By comparison, our approach does not rely on interferometry and is therefore simpler and more robust. Realtime non-perturbing polarization probes can be implemented for other two-level systems (qubits) embedded in the ground manifold, and we have found these useful in a separate project to implement quantum logic in optical lattices.%

\begin{figure*}
%[t]\resizebox{17.5cm}{!}
{\includegraphics{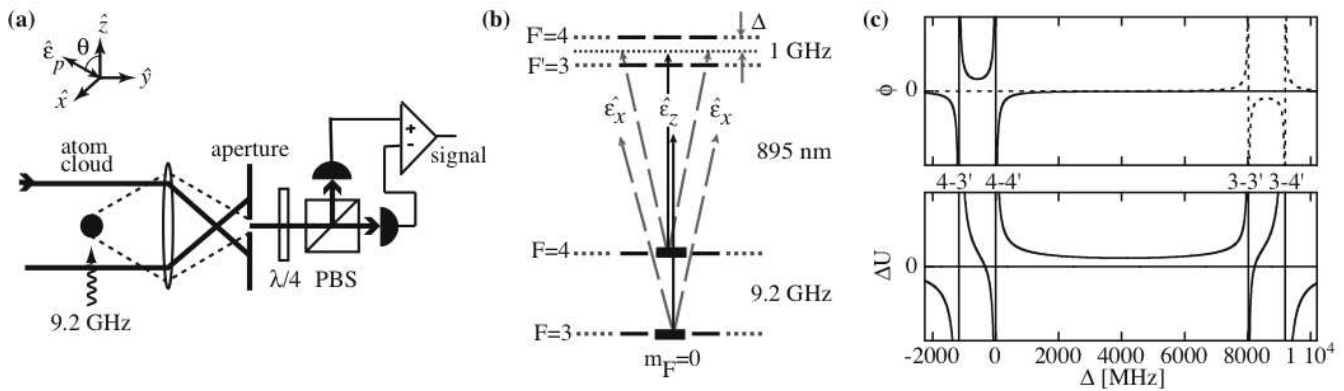}}
\caption{\label{fig:setup} (a) Schematic of our experimental setup. The probe beam propagates along $\hat{y}$, and is linearly polarized at $\theta = 45\,^{\circ} $ in the $\hat{x} - \hat{z}$ plane. (b) Level scheme for the Cs clock states probed on the $\mathrm{D_{1}}$ line, indicating the relevant transitions for the $x$ and $z$ polarized probe components. (c). Top: birefringent phase shift $\phi$ as a function of probe frequency, for atoms in the $|4,0\rangle$ (solid) and $|3,0\rangle$ (dashed) clock states. Bottom: differential light shift $\Delta U$ as a function of the probe detuning from the $F=4 \to F' = 4$ transition, showing the existence of two magic frequencies where $\Delta U = 0$.}
\end{figure*}

	In the far-off-resonance limit the interaction of an alkali atom and a probe field is described by the light shift Hamiltonian, $V_{AP} =\Sigma_{ij} E_{i}^- E_{j}^+ \alpha_{ij}$, where $\alpha_{ij}$ is the ground state polarizability tensor and the tensor $E_{i}^- E_{j}^+ $ contains information about the irradiance and polarization of the probe field \cite{j,l}. In general  $V_{AP}$ has contributions from the scalar, vector and rank-2 tensor components of the polarizability. For probe detunings much larger than the excited state hyperfine splitting, $\Delta \gg \Delta_{HF}$, the scalar and vector components scale as $1/\Delta$, whereas the rank-2 tensor component scales as  $1/\Delta^2$ and is typically ignored. We concentrate at first on a single hyperfine manifold of given $F$, and choose the propagation and quantization axes as in Fig.~\ref{fig:setup}(a). In that case the vector part reduces to $V_{AP}^{(1)} = \xi_{1}J_{3}\tilde{F_{y}}$, where $\mathbf{ \tilde{F}}$ is the collective hyperfine angular momentum and $\mathbf{ J} $ is the Stokes vector for the probe field \cite{k}. It follows that atoms in the $m_{F} = 0$ clock states cannot contribute to Faraday rotation, and therefore in the limit $\Delta \gg \Delta_{HF}$  the probe polarization does not couple to the clock pseudo-spin. For probe detunings $\Delta \sim \Delta_{HF}$, the rank-2 tensor component is significant and can induce a probe ellipticity that depends on the clock pseudo-spin. Here we describe conceptually how such a measurement can be implemented; a more detailed discussion will be published elsewhere \cite{x}. Consider a sample of atoms in the $|F,m_{F}\rangle = |4,0\rangle$  clock state and a probe tuned to the $F=4 \to F'=4$ transition. The $m_{F} = 0 \to m_{F'} = 0$  transition is forbidden and the sample is transparent to the $z$ polarized probe component, while the $m_{F}=0 \to m_{F'} = \pm1$ transitions are allowed and the $x$ polarized component sees a small change in refractive index. The result is a birefringent phase shift $\phi$ of one polarization component relative to the other. When all relevant levels and oscillator strengths are included (Fig.~\ref{fig:setup}(b)) both $x$ and $z$ polarized components undergo non-zero but different phase shifts, and it is straightforward to find the net birefringence for any probe frequency and atoms in either clock state (Fig.~\ref{fig:setup}(c)). Because of the large hyperfine splitting of the $\mathrm{D_{1}}$ multiplet, ($\Delta_{HF}=1168$MHz $=256\Gamma$) we can tune the probe in-between the $F=4 \to F' = 3,4$ transitions to get substantial birefringence with negligible absorption. The probe detuning from the $F=3 \to F' = 3,4$ transitions is roughly the ground hyperfine splitting, ($\sim 2000\Gamma$), and the birefringence due to atoms in the $|3,0\rangle$ state is negligible. Ignoring it we obtain, for an ensemble in a superposition of the two clock states and a probe tuned exactly halfway between the $F=4 \to F' = 3,4$ transitions, a total birefringent phase shift

\begin{equation}
\phi=
\frac{5}{96}\frac{OD}{\Delta/\Gamma}\frac{1}{\tilde{S}}
\left(
\tilde{S_{3}}+\tilde{S}\right)
\label{eq:biref}
\end{equation}

 	Here $OD$ is the optical density of the sample on resonance, and $\Delta=-\Delta_{HF}/2$  is the detuning from the $F=4 \to F'=4$ transition. In addition, we have labeled the $|4,0\rangle,|3,0\rangle$ states as spin-up/down respectively, and introduced the collective pseudo-spin $\mathbf{\tilde{S}} = \sum_{i=1}^{N} \boldsymbol{ \sigma}^{(i)}$, $\tilde{S} = N$ for a sample of $N$ atoms. This result can be compared directly to the benchmark provided by Faraday rotation measurements of a collective angular momentum \cite{faraday}. For a given $OD$ the birefringent phase is $\sim 30\%$ of the Faraday phase that would be seen for an angular momentum with $\tilde{F_{z}}/\tilde{F} = \tilde{S_{3}}/\tilde{S}$. This indicates that our measurement can be expected to resolve projection noise and enter the regime of significant quantum backaction for a cloud $OD$ very similar to that used in \cite{h}.

        A birefringent phase shift $\phi$ corresponds to a rotation of the Stokes vector by an angle $\phi$ around the $\hat{1}$ axis of the Poincar\'{e} sphere. A detailed analysis shows that if we ignore coupling to the spin-down state, the (pseudo)spin-probe interaction is of the simple form $V_{SP} = (\xi_{0} J+\xi_{2} J_{1})(\tilde{S_{3}}+\tilde{S})/2$, where $J$ is the magnitude of the Stokes vector, the constants $\xi_{0}$ and $\xi_{2}$  depend on probe frequency and atomic oscillator strengths, and  $\xi_{0} J+\xi_{2} J_{1}$ is the total (scalar plus tensor) collective light shift for atoms in the upper clock state \cite{x}. In principle $V_{SP}$ has the form required for a QND measurement \cite{k}. To design a measurement that is useful in practice, however, we need to consider two additional aspects of the atom-probe interaction. First, it is essential to minimize the differential light shift $\Delta U$  between the clock states. If we do not, then in a realistic experiment  $\Delta U$ will vary with probe irradiance across the atomic sample, and lead to inhomogeneous broadening of the transition frequency. Fortunately the differential light shift can be eliminated if the probe is tuned to one of two ``magic'' frequencies in-between the $F=3 \to F'$  or  $F=4 \to F'$ transitions (Fig.~\ref{fig:setup}(c)). Second, we must minimize leakage from the clock states to the rest of the ground manifold. Because the probe polarization is linear this involves only the rank-2 tensor component of the light shift, which in a sufficiently strong bias magnetic field along $\hat{z}$  is effectively proportional to $\tilde{F_{z}}^{2}$, and therefore does not couple the clock states to other magnetic sublevels\cite{nonlinear}. Thus, for the correct probe frequency and bias magnetic field, our measurement of $\tilde{S_{3}}$ is non-perturbing, at least on timescales short compared to optical pumping.

\begin{figure}
[t]\resizebox{8.75cm}{!}
{\includegraphics{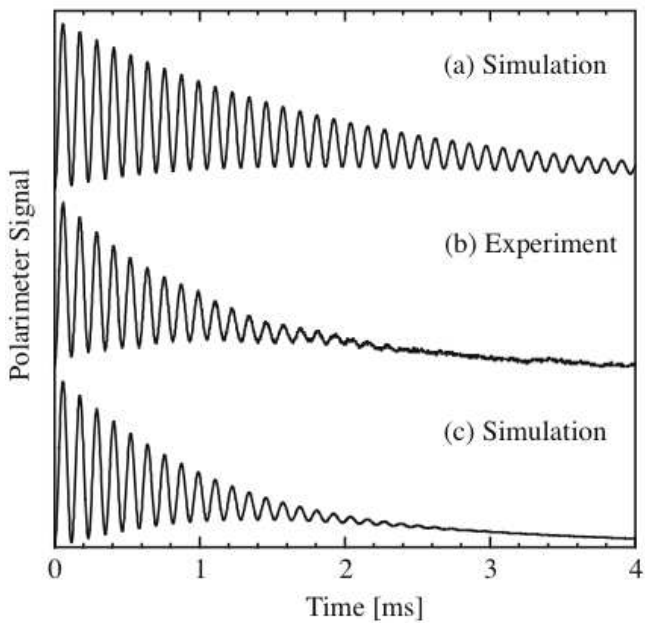}}
\caption{\label{fig:rawdata} (a) Master equation simulation and (b) average of 32 experimental measurement records showing the polarimeter signal due to Rabi oscillations, for a probe tuned to the magic frequency in between the $F=4 \to F'$ transitions, and with an irradiance  $I\sim 16 I_{sat}$. (c) includes the effect of an inhomogeneous microwave field and an additional atom loss rate unrelated to the probe.  }
\end{figure}

The basic setup of our experiment is shown in Fig.~\ref{fig:setup}(a). We begin by preparing a cloud of $3.5\times10^6$ laser cooled atoms, with a 1/e radius of  $\sim0.25$ mm and an estimated $OD=1.8\pm0.5$. After cooling, the atoms are optically pumped into the $|3,0\rangle$ state, released into free fall, and probed on the $F=4\to F'$  transitions.  Our probe beam is generated using a diode laser, spatially filtered with an optical fiber and then passed through a high quality Glan-Laser polarizer before it is incident on the atomic sample. The probe direction is vertical to minimize the effects of atoms falling during the measurement. Its irradiance profile is nearly Gaussian with a $1/e$ radius of ~1.2 mm, which is much larger than the 0.25 mm radius of the atomic cloud, and ensures that the probe light shift is reasonably uniform. We use an imaging system to select only the part of the probe that passes through the cloud, and measure the birefringence with a simple polarimeter consisting of a quarter-wave plate, polarization beamsplitter and differential photo-detector.

	To evaluate the performance of our measurement we drive Rabi oscillations between the clock states with a resonant microwave field at $\sim9.2$GHz\frenchspacing. We apply a $\sim 500$mG magnetic field along the $z$-axis, enough to shift transitions between magnetic sublevels other than the clock states out of resonance with the driving field. This field also prevents the tensor light shift from coupling the clock states to the rest of the ground manifold. Fig.~\ref{fig:rawdata}(b) shows a typical polarimeter signal from Rabi oscillating atoms, for a probe beam tuned to the magic frequency. For comparison, Fig.~\ref{fig:rawdata}(a) shows the calculated signal based on a solution on the master equation, including the microwave drive, probe-induced optical pumping within the entire ground hyperfine manifold, and the sample birefringence caused by atoms in all hyperfine magnetic sublevels including the clock states.  The only free parameters are the microwave Rabi frequency and the cloud optical density.  By matching the signal amplitude at $t = 0$ we infer an $OD = 2.2$, in good agreement with our estimate based on atom number and cloud size.  This simulation shows that the signal decay is fundamentally limited by optical pumping, and that the useful measurement window is therefore inversely proportional to the probe scattering rate.  A better correspondence at later times is obtained in Fig.~\ref{fig:rawdata}(c), where the simulation includes a $1.5 \% $ spatial inhomogeneity in the microwave irradiance, and an additional loss of population from the clock states at a rate of $(2.5ms)^{-1}$. The microwave inhomogeneity has been estimated by seeking the best overall fit to many signals acquired across a wide range of parameters, while the extra loss rate has been independently determined from the asymptotic signal decay rate as the probe scattering rate goes to zero \cite{faraday}. There is no reason to believe the loss is fundamental to our measurement scheme, and we suspect it stems from background gas collisions, stray light or similar problems.  This is supported by the fact that related measurements in a different apparatus have shown loss rates below $(7.5ms)^{-1}$. In practice one can usually increase the probe scattering rate and thus shorten the measurement window to the point where extrinsic losses are unimportant, without compromising the performance of the measurement \cite{faraday}.

\begin{figure}
[t]\resizebox{8.75cm}{!}
{\includegraphics{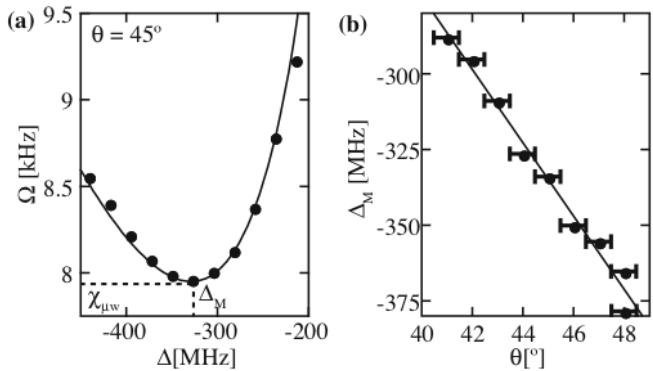}}
\caption{\label{fig:lshift} Signatures of the magic probe frequency. (a) Rabi frequency as a function of probe detuning for a polarization angle $\theta = 45\,^{\circ}$. (b) Magic probe detuning as a function of $\theta$. The solid lines are theory predictions. }
\end{figure}

The non-perturbing aspects of the atom-probe interaction can be examined by tuning the probe in the neighborhood of the magic frequency. Fig.~\ref{fig:lshift}(a) shows the microwave Rabi frequency versus probe frequency. At the magic frequency ($\Delta_{M}=-335$MHz) the Rabi frequency is minimum, $\Omega=\chi_{\mu w}$, corresponding to resonant excitation. At other probe frequencies the differential light shift leads to an effective detuning $\delta_{\mu w}$ between the microwave and light shifted transition frequency, resulting in an increased Rabi frequency  $\Omega = \sqrt{\chi_{\mu w}^{2}+\delta_{\mu w}^{2}}$. Both the increase in $\Omega$ versus probe frequency, and the gradual change in the magic frequency as the probe polarization is rotated (Fig.~\ref{fig:lshift}(b)) agree well with theory, showing that we understand how to detect and eliminate the differential light shift.

\begin{figure}
[t]\resizebox{8.75cm}{!}
{\includegraphics{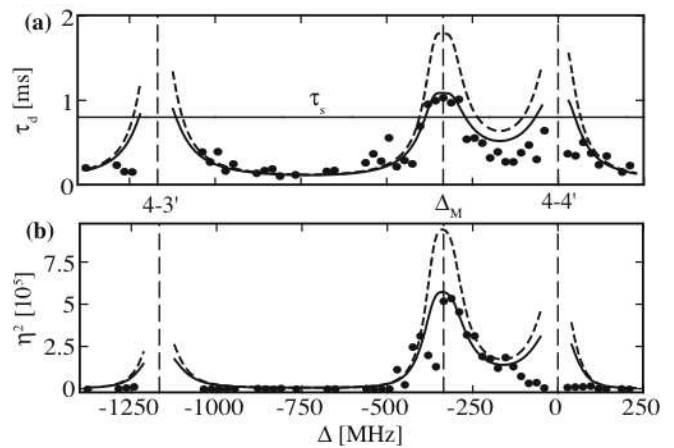}}
\caption{\label{fig:decay} (a) Signal decay $\tau_{d}$ and (b) integrated measurement strength $\eta^{2}$ as a function of probe frequency, for a constant scattering rate of  $(0.8ms)^{-1}$ (assuming equal population of the clock states, on average). The dashed lines assume a probe irradiance inhomogeneity of $\sim15\%$ (rms). The solid lines include an additional atom loss rate as mentioned in the text. Vertical lines indicate the $F=4 \to F'$ transitions and magic probe frequency.}
\end{figure}

The overall performance of our measurement can be quantified by the maximum information gain, i.e. by the measurement strength $\kappa$ integrated over the available measurement time. Fig.~\ref{fig:decay}(a) shows the decay time $\tau_{d}$ for the Rabi oscillations as a function of probe frequency, for constant scattering and  optical pumping rates. Away from the magic frequency the inhomogeneous light shift broadens the distribution of detunings $\delta_{\mu w}$, and cause the Rabi oscillations for different atoms to dephase. The resulting decrease in $\tau_{d}$ is clearly visible in the data. Also shown are predictions based on simulations such as that in Fig.~\ref{fig:rawdata}(c), but now including a Gaussian distribution of probe irradiances with a standard deviation of $\sim15\%$ (best fit). Agreement is generally excellent, except near the atomic resonances where our model fails to account for absorption. Using $\tau_{d}$ as an estimate for the available measurement time, we can approximate the integrated measurement strength by $\kappa\tau_{d}\propto \eta^2$, where $\eta$ is the SNR for a measurement with $\tilde{S_3} = \tilde{S}$ and bandwidth $(\tau_d)^{-1}$. Fig.~\ref{fig:decay}(b) shows values for $\eta^2$ extrapolated from the actual SNR in our Rabi oscillation data, along with the predictions of simulations with $OD=2.5$ (best fit).  Our theory and experimental data agree well and are both strongly peaked at the magic probe frequency.  Finally, a measure for the significance of quantum backaction can be obtained by estimating the SNR in a measurement of the spin projection noise.  Averaging over the entire measurement window $\tau_d$ , this SNR should be $~0.2$ \cite{faraday} for the parameters used in Fig. ~\ref{fig:decay}, which is too low to generate any appreciable degree of spin squeezing.  However, the SNR in a measurement of projection noise scales as $\sqrt{OD}$, so a feasible $OD \sim 10^3$  will increase measurement backaction by a factor of $20$ over our current experiment.  Assuming that we do not encounter additional sources of noise or decoherence when working with dense, correlated samples, an experiment along the lines of \cite{h} might then produce as much as 10 to 15 db of squeezing in the variance of  $\tilde{S_{3}}$\cite{l}. In practice squeezing may be complicated by the fact that our measurement is most sensitive to projection noise near $\tilde{S_{3}} \sim 0$. In contrast to Faraday measurements, our mean signal is then non-zero, $\phi \propto \tilde{S_{3}}+\tilde{S} \propto N$, so that additional noise will arise from fluctuations in the atom number. If necessary this problem can be mitigated by using a probe with two frequency components, tuned between the $F=3 \to F'$ and $F=4 \to F'$ transitions respectively. As seen from Fig.~\ref{fig:setup}(c) the birefringent phase shifts for the probe components have opposite signs and can be arranged to cancel at $\tilde{S_{3}} = 0$. 

In conclusion we have demonstrated a polarimetry-based non-demolition measurement of the pseudo-spin associated with the clock transition in an ensemble of laser cooled Cs atoms. At certain magic frequencies the probe-induced light shift of the clock transition vanishes, and the measurement becomes non-perturbing on time scales shorter than optical pumping.  As a next step we plan to work with samples that are optically thick on resonance, where the sensitivity will be high enough to resolve projection noise. Ultimately we hope to use spin squeezed states to increase sensitivity in Ramsey interrogation of the clock transition, and perhaps improve the performance of atomic clocks and interferometers.

We thank I. H. Deutsch, B. Mischuck, W. Rakreungdet and A. Silberfarb for helpful discussions.  This work was supported by NSF Grants PHY-0099582 and PHY-0355073.

% Comment out next lineafter creating the compiled bbl file and then cut&paste from bbl file

%\bibliography{Clock2005} %Create the reference bibliography via BibTex 

\end{document}